\documentclass[prb,showpacs,twocolumn]{revtex4}
\usepackage{graphicx}

\begin{document}
\title{Electron properties of fluorinated single-layer graphene transistors}
\author{F. Withers$^\dag$}
\author{M. Dubois$^\ddag$}
\author{A. K. Savchenko$^\dag$}
\affiliation{\dag Centre for Graphene Science, School of Physics, University of Exeter, Exeter, EX4 4QL, UK}
\affiliation{\ddag Clermont Universit\'e, UBP, Laboratoire des Mat\'eriaux Inorganiques, CNRS-UMR 6002, 63177 Aubi\`ere, France}

\begin{abstract}
We have fabricated transistor structures using fluorinated single-layer graphene flakes and studied their electronic properties at different temperatures. Compared with pristine graphene, fluorinated graphene has a very large and strongly temperature dependent resistance in the electro-neutrality region. We show that fluorination creates a mobility gap in graphene's spectrum where electron transport takes place via localised electron states.
\end{abstract}

\maketitle

Graphene, a gapless semiconductor with `massless' charge carriers, is a material with many unusual electron properties and potential for new device applications \cite{GeimReview}. However, in a graphene-based transistor the absence of the gap in the band structure results in a relatively small resistance difference between the electro-neutrality (Dirac) region and a region with large carrier concentration (i.e. between the `on' and `off' states). This can become a significant limitation in its use in electronics and intensive research is currently underway aimed at the creation of a (tunable) gap in graphene's energy spectrum. One direction is functionalisation of graphene with suitable elements that will transform its planar crystal structure, with $sp^2$ bonds between the carbon atoms, into a three-dimensional structure with $sp^3$ bonding between them. Theoretical predictions show that hydrogen and fluorine are good candidates to play such a role, with an expected band gap of  3.8 eV and 4.2 eV for 100\% functionalisation, respectively \cite{sofo,Buk1}.

Successful  hydrogenation of mono-layer graphene has already been achieved \cite{KimHydr,GeimHydr}, but there has been no experimental investigation of fluorinated graphene transistors. In this work we have succeeded in fabricating transistor structures with fluorinated graphene monolayers and studied their transport properties at temperatures from 4.2 to 300 K.
Fluorinated graphene flakes were separated by mechanical exfoliation \cite{GeimReview} from fluorinated graphite with fluorine contents of 24\% and 100\%. They are then processed into transistor structures which have shown a strong increase of the resistance in the Dirac region, caused by the opening of a mobility gap in the graphene spectrum.

There are two main ways to produce fluorinated graphite \cite{M1,M2,M4,M6}. In the first graphite is heated in the presence of F$_{2}$ to temperatures in excess of $300^{\circ}$C,
so that covalent C-F bonds are formed and modify the carbon hybridisation. The layered structure of graphite is then transformed into a three-dimensional arrangement of carbon atoms. In the second, graphite is exposed to a fluorinating agent, XeF$_{2}$, and the process is performed at a temperature lower than $120^{\circ}$C as XeF$_{2}$ easily decomposes on the graphite surface into atomic fluorine. Due to its reactivity and diffusion, the fluorination results in a homogenous dispersion of fluorine atoms that become covalently bonded to carbon atoms \cite{M4,M11}. At low fluorine content (F/C atomic ratio $\leq0.4$), conjugated C-C double bonds in the non-fluorinated parts and covalent C-F bonds in corrugated fluorocarbon regions coexist \cite{M6,M7}, with the concentration of the covalent bonds increasing with concentration of fluorine.

In this work we use both methods of fluorination of the original graphite material. We first considered fully fluorinated HOPG graphite (CF)$_{n}$ as a starting material. The flakes are mechanically exfoliated from the fluorinated graphite onto conventional Si/SiO$_{2}$(275nm) substrates. The produced flakes are noticeably smaller ($\sim$1$\mu$m in size) than the flakes fabricated by the same method from non-fluorinated graphite. They  contain more than 10 monolayers - it was impossible to produce mono-layer fluorinated graphene using this method.

\begin{figure}[htb]{}
\includegraphics[width=.85\columnwidth]{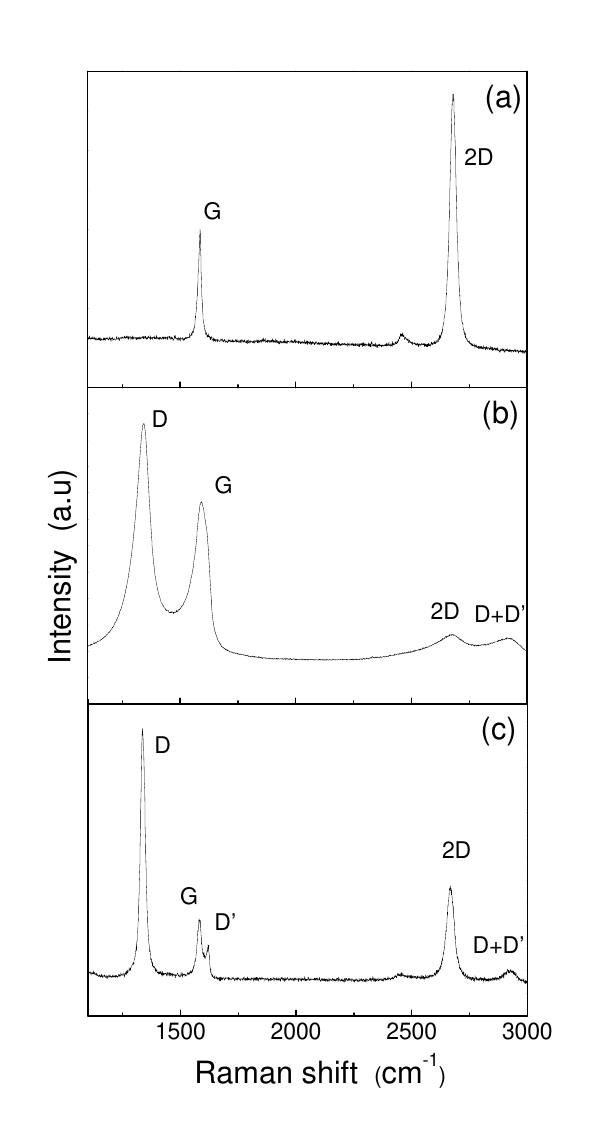}
\caption{\label{fig1} Raman spectra for (a) pristine graphene, (b) multi-layer fully fluorinated graphene and (c) fluorinated graphene monolayer. }
\end{figure}

The many-layer exfoliated flakes were first characterised by Raman spectroscopy using excitation light with a wavelength of 532 nm  and a spot size of 1.5 $\mu$m in diameter. An incident power of $\sim$5 mW was used. We ensured that this power does not damage the graphene by performing Raman measurements on a similarly sized pristine graphene flake which shows
the common spectra of mechanically exfoliated graphene: the G and 2D bands at 1580 cm$^{-1}$ and 2700 cm$^{-1}$, see Figure~\ref{fig1}(a).

In fully fluorinated graphene, a D-peak appears at 1350 cm$^{-1}$ and its intensity is larger than that of the G and 2D peaks, see Figure~\ref{fig1}(b).
As the D resonance requires a defect for its activation, its presence is associated with an increased degree of disorder \cite{R1,R2,R3}. Various defects can contribute to the D-peak, such as bond dislocations, missing atoms at the edges of the sample and sp$^3$ hybridised carbon atoms. 
Previous studies demonstrated that the intensity of the  D-peak produced from the edges of a graphene flake is relatively small compared to the G peak \cite{R4}. Our experiments consistently show that the D-peak of pristine graphene flakes with similar size as fully fluorinated graphene is below the resolution of the measurement, see Figure~\ref{fig1}(a).  
This suggests that the edges of the fully fluorinated flakes give minimal contribution to the D-peak in Figure~\ref{fig1}(b). Futhermore, structural studies \cite{M1} of the bulk fully fluorinated graphite material show that the defects are sp$^3$ bonded carbon atoms and not bond dislocations. Therefore, we expect some contribution to the D-peak of fully fluorinated graphene to come from sp$^3$ bonded carbon atoms.

\begin{figure}[htb]{}
\includegraphics[width=.85\columnwidth]{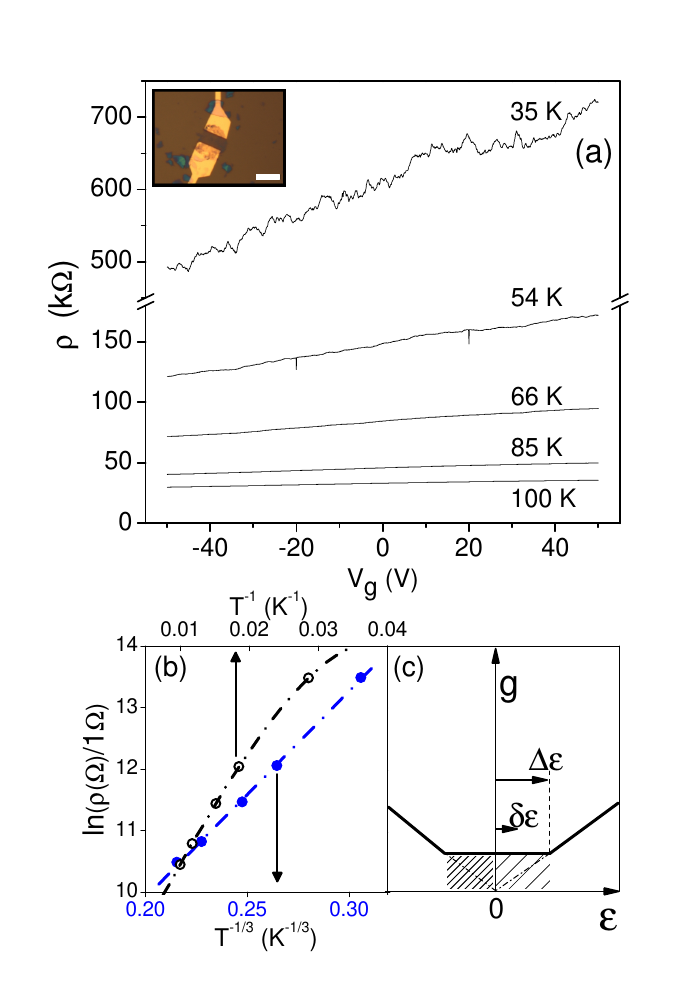}
\caption{\label{fig2} (colour online)  (a) Resistivity as a function of gate voltage for a fully fluorinated device. Inset: Optical image of the device (scale bar is 1 $\mu$m). (b)  Resistivity plotted against $T^{-1}$ and $T^{-1/3}$ at V$_g$=+50 V. (c) A diagram of the energy dependence of the density of electron states, with the Fermi level at zero energy and localised states shown by the shaded area. (Dense shading shows occupied localised states.) }
\end{figure}

Several transistor structures have been fabricated from the fully fluorinated flakes using e-beam lithography and deposition of Cr/Au contacts (5 nm of Cr and 50 nm of Au). All the studied devices show very large resistance (more that 1 GOhm) and no gate-voltage control of the resistance, typical of a wide band gap semiconductor material.
Recent works \cite{f1,f2} have shown that the I/V characteristics of fully fluorinated graphene are strongly non-linear with a nearly gate independent resistance value higher than 1 G$\Omega$, suggesting the presence of a band gap. The purpose of this work is to produce transistor structures with a large on/off ratio of the current. To this end,
we reduced the fluorine content by annealing the samples at $\sim300^{\circ}$C in a $10\%$ atmosphere of H$_{2}$/Ar for 2 hrs. The resistance is decreased and a partial gate-voltage control is achieved, Figure~\ref{fig2}(a). The annealing, however, has not noticeably changed the Raman spectrum in Figure~\ref{fig1}(b).

Resistance measurements of the fully fluorinated flakes after annealing show a strong temperature dependence, Figure~\ref{fig2}(a). To examine the presence of the energy gap,
we analysed $\rho(T)$ at the highest gate voltage $V_g=$50 V, (which is still far from the Dirac point) by an exponential law describing thermal activation of carriers across the energy gap $\Delta\epsilon$:  $\rho=\rho_{0}\exp(\Delta\epsilon/2k_{B}T)$, Figure~\ref{fig2}(b). The resulting value of $\Delta\epsilon$ found at high temperatures is only $\sim$25 meV, which is significantly smaller than the expected energy gap for fully fluorinated graphene. It is also seen that the slope of $\ln\rho(1/T)$ dependence decreases with decreasing $T$, which is a signature of hopping conduction via localised states \cite{hopp}. The fact that in the whole range of studied temperatures electron transport is not due to  thermal activation across the gap but due to hopping, has been confirmed by re-analysing the temperature dependence in terms of two-dimensional hopping: \newline
$\rho(T)=\rho_{0}\exp(T_{0}/T)^{1/3}$, where $k_BT_{0}=13.6/a^{2}g(\mu)$, $g$ is the density of localised states at the Fermi level $\mu$ and $a$ is the localisation length \cite{hopp}. The results are found to be in good agreement with this expression, Figure~\ref{fig2}(b), with the value of $T_{0}=$20000 K. This confirms that the previously found activation energy of 25 meV is not the activation energy $\Delta\epsilon$ that separates the localised states from extended states at the mobility edge, but is an activation energy $\delta\epsilon$ of hopping between localised states within the mobility gap, Figure~\ref{fig2}(c). Although not measured, the value of $\Delta\epsilon$ in these samples seems to be larger than that in hydrogenated graphene, as the obtained value of $T_0$ is $\sim$100 times larger than in \cite{GeimHydr}, indicating a smaller density of localised states and smaller localisation length.

To achieve good $V_g$-control of fluorinated \newline graphene transistors, one needs to fabricate monolayer flakes. To do this, we have used the second method of fluorination, by gaseous XeF$_{2}$. The mixture of natural graphite and XeF$_{2}$ was prepared in a glove box in an Ar atmosphere. The reactor was then kept at $120^{\circ}$C for 48 hrs, which gave a fluorine content of 24\% as measured by mass uptake. Mechanical exfoliation of the partially fluorinated graphite is  carried out under ambient conditions. Flakes are located using an optical microscope and mono-layer flakes with an optical contrast of $\sim5-7\%$ in green light were selected for processing into transistor structures. (We noticed that the contrast of fluorinated monolayers is $\sim2\%$ lower than that of pristine graphene on Si/SiO$_2$(275nm) substrates). The flakes were confirmed to be monolayer by  Raman spectroscopy. The 2D peak which is well fitted  by a single Lorentzian function, and has a FWHM in the region of 20-30 cm$^{-1}$ is typical for pristine monolayer graphene \cite{R1}.

The Raman spectrum of partially fluorinated mono-layer graphene, Figure 1(c), shows much narrower D, G and 2D peaks compared with thicker layers made from fully fluorinated graphite, Figure 1(b). This made it possible to detect additional features in the Raman spectrum that also arise in hydrogenated graphene \cite{GeimHydr}: the $D'$ peak at $\approx 1620$ cm$^{-1}$, which also requires a defect for its activation, and a combination mode  $(D+D')$ at $\approx 2950$ cm$^{-1}$.  It is interesting to note that in our fluorinated graphene the ratio of the integrated intensities $I_{D}/I_{G}=3.8$, which is larger than that found in partially (on one side) hydrogenated sample, and is comparable to that in the samples hydrogenated on both sides \cite{GeimHydr}. This can be the result of the fact that in the monolayer flakes fabricated by exfoliation from fluorinated (intercalated) graphite the probability of fluorination is equal for both sides of the plane.

\begin{figure}[htb]{}
\includegraphics[width=.85\columnwidth]{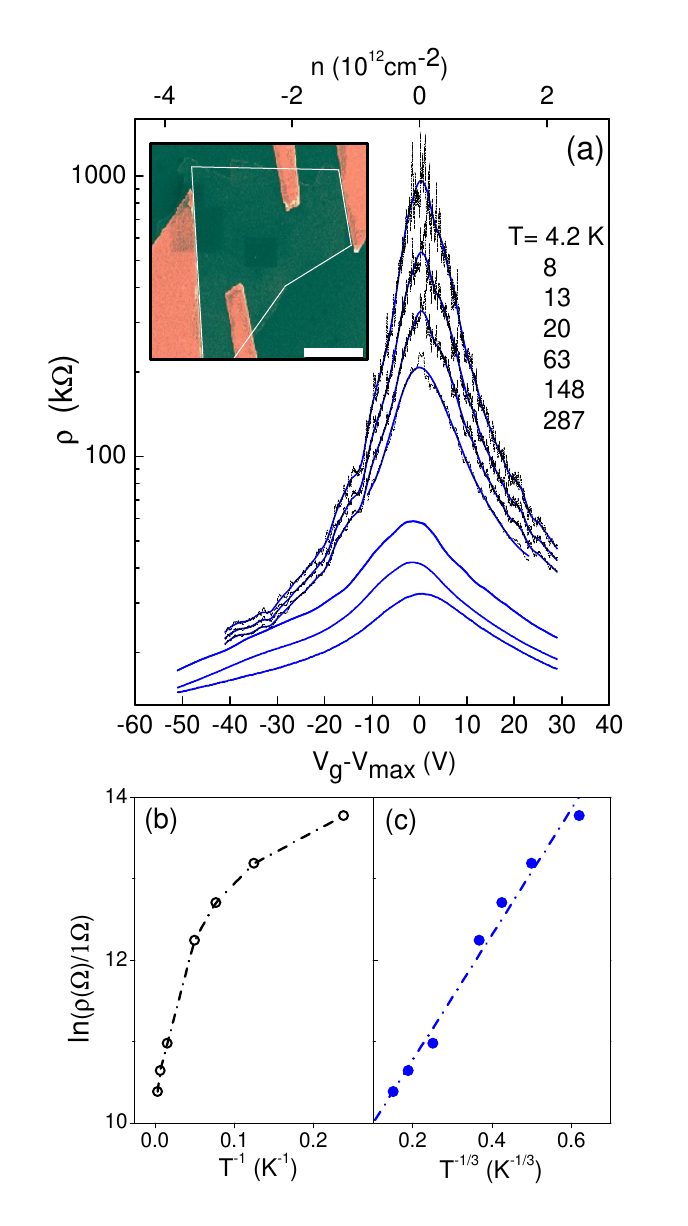}
\caption{\label{fig3} (colour online) (a) Resistivity of fluorinated mono-layer graphene as a function of gate voltage. Inset: false colour SEM image of the mono-layer device (scale bar is 1 $\mu$m).  Resistivity in the Dirac region plotted as a function of (b) $T^{-1}$ and (c) $T^{-1/3}$.}
\end{figure}

In total, four monolayer flakes were processed into four-terminal transistor  devices. Figure~\ref{fig3}(a) shows the resistance as a function of gate voltage measured for a range of temperatures. (Comparisons of two- and four-probe measurements have shown that the contact resistance is negligible compared with the sample resistance, which indicates a good Ohmic contact between the Cr and fluorinated graphene.) Due to the small size of the samples, $\sim4 \mu$m$^{2}$ in area, the resistance shows strong mesoscopic fluctuations \cite{UCF}, and thus the $R(V_g)$ dependences were smoothed for the subsequent analysis using a moving average filter.

The curves in Fig. 3(a) have been offset along the $V_g$-axis to account for doping, which is detected as a shift of the maximum resistance (the Dirac point) from $V_g=0$. The partially fluorinated samples were found to be doped to $V_g=$+10 V ($n=0.74\times 10^{12}$ cm$^{-2}$). This level of doping is similar to that seen
in conventional (pristine) graphene devices, which we attribute to doping by atmospheric water. Unlike pristine graphene devices with very weak temperature dependence of the resistance, in the range of $V_g=$ $\pm 20$ V around the Dirac point the resistivity of fluorinated samples is seen to grow by two orders of magnitude as $T$ decreases from 300 K to 4.2 K. Outside this region the temperature dependence remains weak, with the mobility of carriers $\sim 150$ cm$^{2}$V$^{-1}$s$^{-1}$.

\begin{figure}[htb]{}
\includegraphics[width=.85\columnwidth]{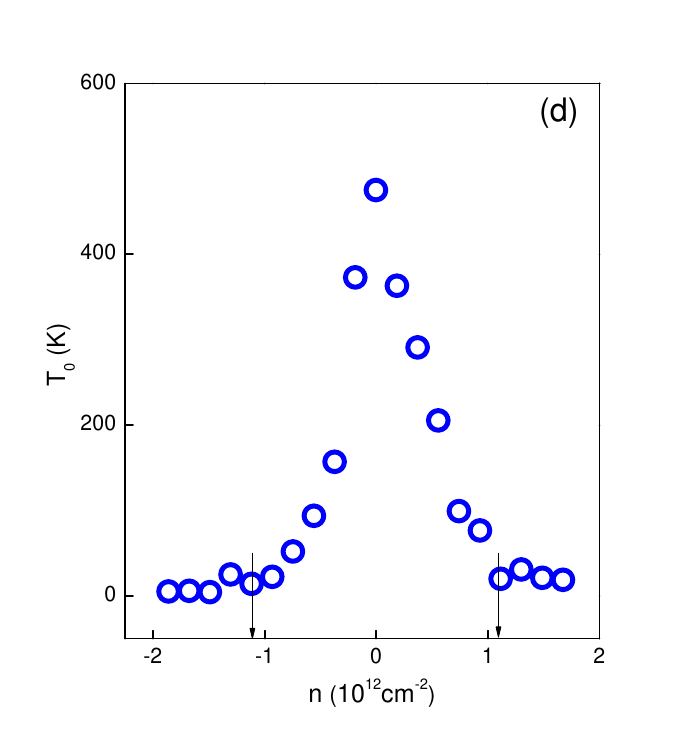}
\caption{\label{fig4} (colour online) (a) The value of the hopping parameter $T_{0}$ as a function of the carrier density. Arrows indicate the concentration at which $T_{0}$ approaches zero and the conduction becomes metallic.}
\end{figure}

Analysis of the temperature dependence of the resistance in terms of the activation law, \newline $\rho=\rho_{0}\exp(\Delta\epsilon/2k_{B}T)$, shows that it is not applicable for the whole temperature range and at high temperatures gives an activation energy of $\sim7$ meV at the neutrality point.
Similarly to the fully fluorinated layers, the resistivity $\rho(T)$ is fitted well by variable range hopping, $\rho(T)=\rho_{0}\exp(T_{0}/T)^{1/3}$, Figure~\ref{fig3}(c). Figure~\ref{fig4} shows the hopping parameter $T_{0}$ as a function of carrier concentration. The value of $T_{0}$ approaches zero at a carrier concentration of $\pm1.2\times10^{12}$ cm$^{-2}$. This value gives the concentration of the localised electron states in the energy range from $\epsilon=0$ to the mobility edge. The mobility edge occurs at $V_g\simeq\pm$20 V and indicates the transition from hopping to metallic conduction, Figure~\ref{fig2}(c).

In order to relate the obtained concentration of the localised states to the energy gap $\Delta\epsilon$, one needs to know the exact energy dependence of the density of states in the gap.  For estimations, we will use the linear relation for the density of extended states above the mobility edge, $g(\epsilon)=2\varepsilon/\pi\hbar^{2}v^2$ ($v=10^6$ ms$^{-1}$ is the Fermi velocity) and a constant value for the density of localised states below the mobility edge, Figure~\ref{fig2}(c). This gives an estimation $\Delta\epsilon\sim60$ meV and twice this value for the full mobility gap. In this approximation the density of the localised states in the gap is estimated as $10^{36}$ J$^{-1}$m$^{-2}$. Using the obtained value of the hopping parameter $T_0\sim$500 K in the Dirac point, one can then estimate the localisation length at $\epsilon=0$ as $a\sim40$ nm.

The presence of localised states in the electro-neutrality region is clearly the result of disorder, due to the random positions of F atoms on graphene in the partial fluorination. The current experiment does not allow us to establish whether these states exist in the band gap produced by fluorination, or are simply the result of a `smearing' of the linear density of states of graphene (as sketched in  Figure~\ref{fig2}(c)). Taking into account the relatively small value of the parameter $T_0$ and large localisation radius $a$, it seems that the latter case is most probable and there is no large band gap created in the spectrum. For future applications, however, it is only an increase of the resistance in the electro-neutrality region which matters, and this is achieved in both scenarios of creation of the mobility gap. Our results show that in order to achieve large resistances in graphene transistors at room temperature, future efforts should be aimed at decreasing the density of localised states in the mobility gap and decreasing the localisation length, which can be done by partial fluorination with the fluorine content in the range between our studied values of 24\% and 100$\%$.

In conclusion, for the first time the possibility of fabricating a transistor structure using fluorinated mono-layer graphene has been demonstrated. Fluorination has been shown to cause a significant increase of the resistance in the electro-neutrality region, which is a consequence of the creation of the mobility gap in the electron spectrum where electron transport is through localised states.\\

\textit{acknowledgments}: We are grateful to A.V.Shytov for useful discussions and A.A.Kozikov for help with measurements, and
to EPSRC for funding.

\end{document}